\documentclass[useAMS,usenatbib,usegraphicx]{mn2e}

\usepackage{amssymb,amsmath}

\title[]{Modelling the VHE flare of 3C\,279 using one zone leptonic model}
\author[S. Sahayanathan and S. Godambe]{S. Sahayanathan$^{1}$\thanks{E-mail:
sunder@barc.gov.in} and S. Godambe$^{2}$\thanks{Email: sagar@physics.utah.edu}\\
$^{1}$Astrophysical Sciences Division, Bhabha Atomic Research Centre, Mumbai - 400085, India \\
$^{2}$Department of Physics and Astronomy,The University of Utah,Salt Lake City, Utah-84112, USA.\\}

\begin{document}

\maketitle

\label{firstpage}

\begin{abstract}
We model the simultaneous observations of the flat spectrum radio quasar 3C\,279 at radio, 
optical, X-ray and very high energy (VHE) gamma ray energies during 2006 flare 
using a simple one zone leptonic model.
We consider synchrotron emission due to cooling of a non-thermal electron distribution in
an equipartition magnetic field and inverse Compton emission due to the scattering off synchrotron
photons (SSC) and external soft photons (EC) by the same distribution of electrons. 
We show that the VHE gamma ray flux cannot be explained by SSC process thereby 
suggesting the EC mechanism as a plausible emission process at this energy. The EC scattering of 
BLR photons to VHE energies will be in Klein-Nishina regime predicting a steep spectrum which is 
contrary to the observations. However the infrared photons from the dusty torus can be boosted to 
VHE energies with the scattering process remaining in the Thomson regime. Though EC process can 
successfully explain the observed VHE flux, it require a magnetic field much lower than the 
equipartition value to reproduce the observed X-ray flux. Hence we attribute the X-ray emission due 
to SSC process. We derive the physical parameters of 3C\,279 considering the above mentioned emission 
processes. In addition we assume the size of emission region constrained by a variability timescale 
of one day. This model can successfully reproduce the broadband spectrum of 3C\,279 but predicts
substantially large flux at MeV-GeV energies.

\end{abstract}


\section{Introduction}

Flat Spectrum Radio Quasars (FSRQ) are radio loud Active Galactic Nuclei (AGN) with 
broad emission lines and a 
non-thermal spectrum extending from radio to gamma ray energies. 
They are characterized by luminous core, rapidly variable non-thermal emission, high 
radio and optical polarization, flat-spectrum radio emission and/or superluminal motion. Similar 
properties are also observed in BL Lacs and these two types
of AGN are classified as blazars. According to AGN unification scheme, blazars are the class
of AGN with a relativistic jet pointed close to the line of sight of the 
observer (\cite{urry95}).  
The spectral energy  distribution (SED) of FSRQ is characterized by two broad humps. The 
low energy hump in the SED is well understood as synchrotron emission from a relativistic 
distribution of electrons. Whereas, the origin of high
energy hump is still matter of debate. 
There are several models to explain the high energy emission
based on either leptonic or hadronic interactions (\cite{BlandfordLevinson95, 
BloomMarscher96, Aharonian2000, PohlSchlickeiser2000, mannheim98}). Under the assumption of
leptonic models, high energy emission is explained via inverse Compton scattering of soft
target photons.  
The target photons can be the synchrotron photons
(synchrotron self Compton (SSC))
(\cite{konigl81,marscher85,ghisellini89}) or the photons external to the jet 
(external Compton (EC))(\cite{begelman87,melia89,dermer92}).
These external photons can be the accretion disk photons (\cite{dermer93,boettcher97}) 
or the accretion disk photons 
reprocessed by broad line region (BLR) clouds
(\cite{sikora94,ghisellini96}) or the infrared radiation from the dusty torus
(\cite{sikora94,blazejowski00,ghisellini08}). 
Hadronic models explain the high energy emission as an outcome of the synchrotron proton 
emission and proton-photon interactions with the synchrotron photons (synchrotron proton blazar model)
(\cite{mannheim98,mucke03,boettcher09}).

3C\,279 ($z=0.536$) is a well studied source at various energy bands with several
simultaneous multi wavelength campaigns(\cite{maraschi94,hartman96,wehrle98,boettcher07}).
It was the first blazar observed at gamma ray 
energies by the satellite based experiment {\it EGRET(Energetic Gamma-Ray Experiment Telescope)}
(\cite{hartman92}) and the first FSRQ to be detected at GeV-TeV (very high energy (VHE))
gamma ray energies by ground based atmospheric Cherenkov experiment {\it MAGIC
(Major Atmospheric Gamma-Ray Imaging Cherenkov)}(\cite{albert08}). 
The flux of 3C\,279 is also known to be strongly variable at radio, 
infrared, optical, UV, X-ray and gamma-ray energies (\cite{makino89} and references therein). 
During 2006, 3C\,279 undergone a dramatic flare and was observed by WEBT (Whole Earth blazar Telescope) campaign 
(\cite{boettcher07}) at radio, near infrared and optical frequencies and simultaneously
monitored by {\it Rossi X-ray Timing Explorer (RXTE)} at X-ray energies (\cite{chatterjee08})
and by {\it MAGIC} telescope at VHE gamma rays (\cite{albert08}).

The SED of 3C\,279 is modelled as synchrotron and inverse Compton emission using these 
simultaneous observations at various energies.
The explanation of the Compton hump as an outcome of SSC process
due to the cooling of a relativistic power-law electron distribution is not very
successful (\cite{boettcher09}). This interpretation 
had a serious drawbacks in explaining the 
high energy gamma ray ($> 1GeV$) detection from 3C\,279 since it require very 
high radiation energy density compared to magnetic energy density(\cite{maraschi92}).
It also require unusually low magnetic field to explain the VHE emission(\cite{boettcher09}). 
However in case of BL Lac objects, the models considering the SSC process are able to 
reproduce gamma ray emission successfully (\cite{stecker96,coppi99,bhattacharyya05}).
Other alternative explanation for the Compton hump of 3C\,279 is the EC mechanism with the 
target photons external to the jet. 
Target photons from the accretion disk will be 
strongly deboosted due to high Lorentz factor of the jet ($\Gamma\sim 10-25$) and 
may not play an important role (however see \cite{boettcher97} for an alternative). 
The photons from BLR clouds cannot be boosted to VHE gamma rays 
since the scattering
of Lyman alpha line emission (the dominant emission from BLR clouds (\cite{francis91})) will 
result in Klein-Nishina regime (\cite{ghisellini09}; section \S \ref{sec:ec}) predicting a 
steep spectrum contrary to the observed hard spectrum. 
The third option, infrared photons from the dusty torus, can be 
a plausible candidate for the target
photons in EC process since the scattering of these photons to VHE gamma ray energies
will still be in Thomson regime (\cite{ghisellini09}; section \S \ref{sec:ec}).
The importance of Compton scattering of infrared photons from dusty torus 
was first studied in detail by \cite{blazejowski00} to explain high energy gamma 
rays detected by {\it EGRET}.
Though EC process is widely accepted to explain the gamma ray emission of 3C\,279, 
\cite{lindfors05} 
argues in favour of SSC mechanism based on observed radio-gamma ray correlation 
and the quadratic dependence of the synchrotron and inverse Compton peak fluxes during the flare. 

In the present paper we use the simultaneous observation of 3C\,279 at radio, near 
infrared, optical, X-ray and very high energy (VHE) gamma ray energies 
(\cite{boettcher07,chatterjee08,albert08})
to deduce the physical parameters of the source. We consider synchrotron, synchrotron 
self Compton (SSC) and external Compton (EC) emission from a power-law distribution 
of electrons. 
We assume a magnetic field which is in 
equipartition with the particle energy. In the next section we outline the model and 
in section \S \ref{sec:ic} we show the plausible target photons 
for the inverse Compton process to
explain the observed flux at different energy bands. In section 
\S \ref{sec:result} we present the estimated physical parameters of 3C\,279 and discuss 
implications of the present model and the results. A cosmology with $\Omega_m = 0.3$,
$\Omega_\Lambda = 0.7$ and $H_0 = 70\;km\;s^{-1}\;Mpc^{-1}$ is used in this work.

\section[]{Model}
We assume the emission region to be a spherical blob moving down the jet at 
relativistic speed with Lorentz factor $\Gamma$. Since the jets of blazars are aligned towards the
line of sight of the observer, we assume the Doppler factor $\delta \approx \Gamma$.
The radiation from emission region is due to synchrotron and inverse Compton cooling of 
relativistic electrons described by a broken power-law distribution
(quantities with prime are measured at the rest frame of the emission region)
\begin{equation}
\label{eq:broken}
N^\prime(\gamma^\prime) d\gamma^\prime =\left\{
\begin{array}{ll}
K \gamma^{\prime-p}d\gamma^\prime,&\mbox {~$\gamma^\prime_{min}<\gamma^\prime<\gamma^\prime_b$~} \\
K \gamma^{\prime(q-p)}_b \gamma^{\prime-q}d\gamma^\prime,&\mbox {~$\gamma^\prime_b<\gamma^\prime<\gamma^\prime_{max}$~}
\end{array}
\right.
\end{equation}
where $\gamma^\prime$ is the Lorentz factor of the electron, $p$ and $q$ are the power-law
indices before and after the break corresponding to the Lorentz factor $\gamma_b^\prime$ and
$K$ is the particle normalisation.
In the blob frame protons are assumed to be cold and contribute only to the inertia of the jet. 
The emission region is permeated with tangled magnetic field $B$. If we assume the 
equipartition condition between the magnetic field energy density($U_B$) and
the particle energy density, then the equipartition magnetic field $B_{eq}$ will satisfy
\begin{equation}
\label{eq:equipart}
U_B = \frac{B_{eq}^2}{8\pi}= mc^2\int\limits_{\gamma^\prime_{min}}^{\gamma^\prime_{max}} \gamma^\prime N^\prime(\gamma^\prime)d\gamma^\prime
\end{equation}
where $m$ is the electron rest mass and $c$ is the velocity of light. We neglect the 
contribution of protons in equation (\ref{eq:equipart}) due to the assumption they are 
cold and do not contribute to the particle energy density.
The size of the emission region $R^\prime$ can be approximated from the observed variability 
timescale $t_{var}$ as
\begin{equation}
\label{eq:size}
R^\prime \approx \frac{\delta}{(1+z)}ct_{var}
\end{equation}
where $z$ is the redshift of the source.  

\section[]{Inverse Compton Process in 3C\,279}{\label{sec:ic}}
The radio-to-optical emission of 3C\,279 is considered to be synchrotron emission 
and X-ray-to-$\gamma$-ray emission is produced by inverse Compton scattering of  
soft photons. The soft target photons can be the synchrotron photon itself (SSC) and/or 
the photon field external to the jet (EC). 

\subsection[]{Synchrotron Self Compton (SSC)}{\label{sec:ssc}}

In SSC mechanism, the synchrotron photons are boosted to high energies via inverse Compton scattering 
by the same electron distribution responsible for the synchrotron emission itself.
An approximate expression for synchrotron emissivity $\epsilon^\prime_{syn}(\nu^\prime) 
[ergs\;cm^{-3} s^{-1} Sr^{-1} Hz^{-1}]$ due to an
electron distribution $N^\prime(\gamma^\prime)$ can be obtained by describing the single particle emissivity in 
terms of a delta function $\delta(\nu^\prime - \gamma^{\prime 2} \nu_L)$ (\cite{shu91})
(Appendix \ref{appendix:syn})
\begin{equation}
  \epsilon^\prime_{syn}(\nu^\prime)\approx\frac{\sigma_TcB^2}{48 \pi^2}\nu_L^{-\frac{3}{2}}
  N\left(\sqrt{\frac{\nu^\prime}{\nu_L}}\right)\nu^{\prime{\frac{1}{2}}}
\end{equation}
Where $\sigma_T$ is Thomson cross section, $\nu_L=\frac{eB}{2\pi mc}$ is the Larmor frequency
and $\nu^{\prime}$ is the frequency of the emitted photon in 
the rest frame of the emission region.
The observed flux $F_{syn}(\nu)[ergs\;cm^{-2}s^{-1}Hz^{-1}]$ will be(\cite{begelman84})
\begin{equation}
\label{eq:flux}
F_{syn}(\nu)\approx \frac{\delta^3(1+z)}{d_L^2} V^\prime \epsilon^\prime_{syn}\left(\frac{(1+z)}{\delta}\nu
\right)
\end{equation}
Where $V^\prime$ is the volume of the emission region and $d_L$ is the luminosity distance. 
The rising part of the observed synchrotron flux in the SED due to the particle distribution 
given by equation(\ref{eq:broken}) can then be written as
\begin{equation}
\label{eq:synflx}
F_{syn}(\nu) \approx s(z,p) \delta^{\frac{p+5}{2}}B^{\frac{p+1}{2}}
R^{\prime3}K\nu^{-\frac{p-1}{2}} \; Jy
\end{equation}
for $\nu < \delta\gamma^{\prime 2}_b\nu_L/(1+z)$. Here $s(z,p)$ is a function of $p$ and $z$. 
For $z= 0.536$ and $p = 2.02$ (corresponding to a radio spectral index $0.51$), 
$s = 1.7\times 10^{-52}$.

The observed synchrotron and SSC frequency corresponding to the peak flux in the SED 
can be approximated as  
\begin{align}
\label{eq:synpeak}
\nu_{p,syn} \approx \frac{\delta}{(1+z)}\gamma_b^{\prime 2} \nu_L  \\
\label{eq:sscpeak}
\nu_{p,ssc} \approx \frac{\delta}{(1+z)}\gamma_b^{\prime 4} \nu_L  
\end{align}
From equations(\ref{eq:synpeak}) and (\ref{eq:sscpeak}) we get 
\begin{equation}
\label{eq:peakenergy}
\gamma^\prime_b = \sqrt{\frac{\nu_{p,ssc}}{\nu_{p,syn}}}
\end{equation}
Using equations(\ref{eq:equipart}), (\ref{eq:size}), (\ref{eq:synflx}) and (\ref{eq:peakenergy}) 
(for $p=2.02$) we get
\begin{align}
\label{eq:gamapeak}
\nu_{p,ssc}&\sim 1.53\times 10^{19}\left(\frac{F_{syn}(\nu)}{8 Jy}\right)^{-0.28}
\left[
\left(\frac{t_{var}}{1day}\right)\left(\frac{\delta}{26}\right)\right]^{0.85} 
\nonumber \\
&\times
\left(\frac{\nu_{p,syn}}{2.2\times10^{13} Hz}\right)^2  
\left(\frac{\gamma^\prime_{min}}{40}\right)^{0.006} 
\left(\frac{\nu}{3\times10^{11} Hz}\right)^{-0.51} Hz
\end{align}
However the gamma-ray peak frequency expressed in equation(\ref{eq:gamapeak}) is too low to 
interpret the observed VHE flux as an outcome of SSC mechanism 
(see Fig(\ref{fig:fit})).

The emissivity due to SSC mechanism 
$\epsilon^\prime_{ssc}(\nu^\prime)[ergs\;cm^{-3} s^{-1} Sr^{-1} Hz^{-1}]$ 
for the particle distribution given by equation(\ref{eq:broken}) can be approximated as 
(Appendix \ref{appendix:ssc})

\begin{equation}
\label{eq:sscemiss}
\epsilon^\prime_{ssc}(\nu^\prime) \approx \frac{R^\prime c}{36\pi^2}K^2\sigma_T^2B^2\nu_L^{-\frac{3}{2}}
\nu^{\prime\frac{1}{2}}\mathnormal{f}(\nu^\prime)
\end{equation}
The function $\mathnormal{f}(\nu^\prime)$ is given by
\begin{align}
\label{eq:f_ssc}
\mathnormal{f}(\nu^\prime) = \Bigg[\left(\frac{\nu^\prime}{\nu_L}\right)^{-\frac{p}{2}}
ln\left(\frac{\gamma^\prime_1}{\gamma^\prime_2}\right) 
+\frac{\gamma_b^{\prime (q-p)}}{q-p}\left(\frac{\nu^\prime}{\nu_L}\right)^{-\frac{q}{2}}
\nonumber \\ \times
(\gamma_1^{\prime (q-p)}-\gamma_{min}^{\prime (q-p)})
\Theta\left(\frac{1}{\gamma^\prime_b}\sqrt{\frac{\nu^{\prime}}{\nu_L}}-\gamma^\prime_{min}\right)
\Bigg]\Theta(\gamma^\prime_2-\gamma^\prime_1)
\nonumber \\ 
+\Bigg[\gamma_b^{\prime 2(q-p)}\left(\frac{\nu^\prime}{\nu_L}\right)^{-\frac{q}{2}}
ln\left(\frac{\gamma^\prime_4}{\gamma^\prime_3}\right) +
\frac{\gamma_b^{\prime (q-p)}}{q-p}\left(\frac{\nu^\prime}{\nu_L}\right)^{-\frac{p}{2}}
\nonumber \\ \times
(\gamma_4^{\prime (p-q)}-\gamma_{max}^{\prime (p-q)})
\Theta\left(\gamma^\prime_{max}-\frac{1}{\gamma^\prime_b}\sqrt{\frac{\nu^{\prime}}{\nu_L}}\right)\Bigg]
\Theta(\gamma^\prime_4-\gamma^\prime_3)
\end{align}
where $\Theta$ is the Heaviside function and 
\begin{align}
\gamma^\prime_1 = MAX\left(\gamma^\prime_{min},\frac{1}{\gamma^\prime_b}\sqrt{\frac{\nu^{\prime}}{\nu_L}}\right) \\
\gamma^\prime_2 = MIN\left(\gamma^\prime_b,\frac{1}{\gamma^\prime_{min}}\sqrt{\frac{\nu^{\prime}}{\nu_L}}\right) \\
\gamma^\prime_3 = MAX\left(\gamma^\prime_b,\frac{1}{\gamma^\prime_{max}}\sqrt{\frac{\nu^{\prime}}{\nu_L}}\right) \\
\label{eq:ssc_g2}
\gamma^\prime_2 = MIN\left(\gamma^\prime_{max},\frac{1}{\gamma^\prime_b}\sqrt{\frac{\nu^{\prime}}{\nu_L}}\right) 
\end{align}
The observed SSC flux 
can be obtained following equation(\ref{eq:flux}) by replacing $\epsilon^\prime_{syn}(\nu^\prime)$ 
with $\epsilon^\prime_{ssc}(\nu^\prime)$. It should be noted here in deriving 
equations (\ref{eq:f_ssc})- (\ref{eq:ssc_g2}) we have assumed the scattering in Thomson regime. 
However at VHE energies this may not be valid since Klein-Nishina effect will be
more prominent and should be included (\cite{tavecchio98}).

\subsection[]{External Compton (EC)}{\label{sec:ec}}
 
In EC mechanism the target photons for the inverse Compton scattering are accretion disk
photons,  
reprocessed disk photons
by the BLR clouds and infrared photons from the dusty torus. In the following discussion 
we consider photons from the BLR clouds and the dusty torus only, since the photons from the 
accretion disk will be strongly redshifted. 

If the VHE emission is produced due to inverse Compton scattering happening in Klein-Nishina regime, 
then the VHE spectrum should be steeper than the observed synchrotron spectrum. However the 
intrinsic VHE spectrum of 3C\,279 is  substantially harder after correction for the absorption 
effects due to extra galactic background light (EBL) (\cite{albert08}).
(Though the considered EBL model (\cite{primack05}) is debatable (\cite{stecker09}).
Therefore, from the inferred intrinsic VHE spectrum, one can conclude that the scattering 
process must be in Thomson regime. 

\cite{ghisellini09} showed that the scattering of BLR photons to VHE energies 
will be in Klein-Nishina regime. The condition for Thomson scattering is given by
\begin{equation}
\label{eq:thomsoncond}
\frac{\gamma \Gamma h \nu_t^\star}{m c^2}<1
\end{equation} 
Where $h$ is the Planck constant and $\nu_t^\star$ is the frequency of target photon in AGN 
frame (quantities with star are measured at the AGN frame). The frequency of the 
scattered photon in the observer's frame is
\begin{equation}
\label{eq:thomsonphot}
\nu \approx \frac{\Gamma^2 \gamma^2 \nu_t^\star}{(1+z)}
\end{equation}
From equations(\ref{eq:thomsoncond}) and (\ref{eq:thomsonphot}) we get
\begin{equation}
\label{eq:thomsontarg}
\nu_t^\star<\frac{1}{\nu(1+z)}\left(\frac{m c^2}{h}\right)^2\sim 10^{14}
\left(\frac{\nu}{10^{26}}\right)^{-1}Hz 
\end{equation}
The dominant emission from BLR region is the line emission at $2.47 \times 10^{15}$ Hz 
corresponding to Lyman-$\alpha$ (\cite{francis91}) and hence the scattering will be in 
Klein-Nishina regime. However the scattering of infrared photon ($10^{12}-10^{14}$ Hz)
from the dusty torus to 
VHE energies can still remain in Thomson regime. 
The interpretation of VHE emission
as a result of inverse Compton scattering of infrared photons is also used to explain
the VHE emission from the intermediate BL Lac objects, W\,Comae and 3C\,66A, detected by 
{\it VERITAS} (\cite{acciari08,abdo11}).

The EC emissivity $\epsilon^\prime_{ec}(\nu^\prime)[ergs\;cm^{-3} s^{-1} Sr^{-1} Hz^{-1}]$ due to 
scattering of an isotropic mono energetic 
photon field of frequency $\nu^\star$ and energy density $u^\star$ (in lab frame) can be 
approximated as (\cite{dermer95})

\begin{equation}
\label{eq:ecemiss}
\epsilon^\prime_{ec}(\nu^\prime)\approx\frac{c\sigma_Tu^\star}{8\pi\nu^\star}
\left(\frac{\Gamma \nu^\prime}{\nu^\star}\right)^{\frac{1}{2}} 
N^\prime\left[\left(\frac{\nu^\prime}{\Gamma\nu^\star}\right)^{\frac{1}{2}}\right]
\end{equation}
The observed EC flux can be obtained following equation(\ref{eq:flux}) by replacing 
$\epsilon^\prime_{syn}(\nu^\prime)$ with $\epsilon^\prime_{ec}(\nu^\prime)$.

If the observed Compton flux can be represented by a broken power-law
\begin{equation}
F_{ec}(\nu) \propto \left\{
\begin{array}{ll}
 \nu^{-\alpha},&\mbox {~$\nu<\nu_{p,ec}$~} \\
 \nu_{p,ec}^{(\beta-\alpha)} \nu^{-\beta},&\mbox {~$\nu>\nu_{p,ec}$~}
\end{array}
\right.
\end{equation}
with $\alpha=\frac{p-1}{2}$ and $\beta=\frac{q-1}{2}$ then the peak Compton
frequency in SED will be
\begin{equation}
\label{eq:comppeak}
\nu_{p,ec} = \left(\frac{F_{ec}(\nu_1)\nu_1^\alpha}{F_{ec}(\nu_2)\nu_2^\beta}\right)^{\frac{1}{\alpha-\beta}}
\end{equation} 
where $\nu_1<\nu_{p,ec}$ and $\nu_2>\nu_{p,ec}$.
Since $\nu_{p,ec}$ corresponds to the scattering off the target photon by the particle of
energy $\gamma^\prime_b mc^2$, we can write
\begin{equation}
\label{eq:ecpeak}
\nu_{p,ec} \approx \frac{\Gamma^2}{(1+z)}\gamma_b^{\prime 2} \nu^\star  \\
\end{equation}
If we assume both the X-ray and the VHE emission are produced by the EC scattering off infrared photons 
itself, then
using equations(\ref{eq:synpeak}), (\ref{eq:comppeak}) and (\ref{eq:ecpeak}), we can write 
\begin{align}
\label{eq:ecequi1}
B &\sim 5.4\times 10^{-2}\left(\frac{\Gamma}{26}\right)
\left(\frac{\nu^\star}{5 \times 10^{13}}\right)
\left(\frac{\nu_{p,syn}}{2.2 \times 10^{13}}\right)\nonumber \\
&\times \left(\frac{(F_{ec}(\nu_1)/1.5\times10^{-6}Jy)(\nu_1/10^{18})}
{(F_{ec}(\nu_2)/1.7\times10^{-12}Jy)(\nu_2/2.1\times10^{25})}\right) \; G
\end{align}
Here we have used $\alpha = 0.51$ and $\beta = 1.6$ corresponding to observed spectral
indices at radio/X-ray and optical energies.
However the equipartition magnetic field deduced from equations(\ref{eq:equipart}), 
(\ref{eq:size}) and (\ref{eq:synflx}) is 
\begin{align}
\label{eq:ecequi2}
B_{eq}&\sim 0.67\left(\frac{\Gamma}{26}\right)^{-1.85}\left(\frac{F_{syn}(\nu)}{8 Jy}\right)^{0.28}
\left(\frac{t_{var}}{1day}\right)^{-0.85}
\nonumber \\
&\times 
\left(\frac{\gamma^\prime_{min}}{40}\right)^{-0.006}
\left(\frac{\nu}{3\times10^{11} Hz}\right)^{0.15} G
\end{align}
Hence we require a magnetic field much lower than its equipartition value to reproduce
the observed X-ray flux through EC scattering of the infrared photons.
However if X-ray emission is due to SSC process, the required magnetic field 
will be close to its equipartition value.

Based on these arguments, we model the broadband spectrum of 3C\,279 considering 
synchrotron, SSC and EC processes. The synchrotron emission is dominant at radio/optical
energies whereas SSC is dominant at X-ray energy and EC at VHE gamma ray energy.

\section{Result and Discussion}
\begin{figure}
\includegraphics[height=0.35\textheight, width=0.35\textwidth, angle = -90] {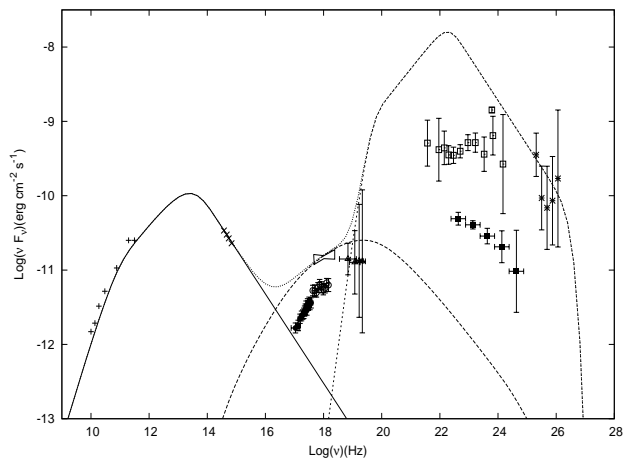}
\caption{ Spectral fit to the SED of 3C\,279. The radio(plus), optical(cross) and {\emph RXTE} 
X-ray data (butterfly) are 
obtained from Boettcher et al. (2007) and the VHE data is reproduced from Albert et al. (2008). 
The {\emph Chandra} (filled circle) and {\emph Swift} (open circle) X-ray data and the 
{\emph INTEGRAL} (triangle) data are obtained from Collmar et al. (2010).
The MeV-GeV data represented by open boxes are the {\emph EGRET} data obtained from Hartman et al. (2001). 
The MeV-GeV data represented by filled boxes are {\emph Fermi} data obtained during 2008 (Abdo et al. (2010)). 
The dashed curve is the synchrotron spectrum, the 
dotted curve is the SSC spectrum and the dot-dashed curve is the EC spectrum. The solid
curve is the total contribution of the different emission processes.}\label{fig:fit}  
\end{figure}

{\label{sec:result}}
The
main parameters governing the observed broadband spectrum of 3C\,279 are the $B$, $K$, $p$, $q$,
$\gamma^\prime_b$, $R^\prime$, $\Gamma$, $\nu^\star$ and $u^\star$. Out of these the indices $p$ and $q$
can be constrained using the observed spectral indices at radio and optical energies. 
If we assume the target photon energy density $u^\star$ from the dusty torus as a blackbody, then the 
peak target photon frequency $\nu^\star$ can be written in terms of $u^\star$ as
\begin{equation}
\nu^\star = 2.82 \frac{K_B}{h}\left(\frac{4\sigma}{c}\int u^\star d\nu^\star\right)^{-\frac{1}{4}}
\end{equation}
where $K_B$ is the Boltzmann constant and $\sigma$ is the Stefan-Boltzmann constant.  
The remaining six parameters $B$, $K$, $\gamma^\prime_b$, $R^\prime$, $\Gamma$ and $u^\star$
can be calculated using equations(\ref{eq:equipart}), (\ref{eq:size}), (\ref{eq:synflx}),
(\ref{eq:synpeak}), (\ref{eq:sscemiss}) and (\ref{eq:ecemiss}). We consider one day as the 
variability timescale $t_{var}$ (\cite{hartman96}). The derived parameters are given in 
table (\ref{tab:parameters}) and the corresponding broadband spectrum is shown in 
figure (\ref{fig:fit}). 
\begin{table}
\label{tab:parameters}
\caption{Model Parameters}
\begin{tabular}{lcc}
\hline
Parameter & Symbol & Numerical Value \\
\hline
Particle spectral index  & $p$ & $2.02$\\
{\it(low energy)} & & \\
Particle spectral index  & $q$ & $4.2$\\
{\it(high energy)} & & \\
Magnetic Field & $B_{eq}$ & $0.67\; G$\\
{\it(equipartition)} & & \\
&&\\
Particle energy density & $U_e$ & $1.8\times 10^{-2}\; ergs/cm^3$\\
&&\\
Particle spectrum break energy & $\gamma_b^\prime$ & $832$\\
{\it(units of $mc^2$)}&&\\
Emission region size&$R^\prime$&$4 \times 10^{16}\;cm$\\  
&&\\
Bulk Lorentz factor & $\Gamma$ & $26$\\  
&&\\
IR dust temperature & $T^\star$ & $860\;K$\\
&&\\
IR dust photon energy density & $u^\star$ & $4 \times 10^{-3}\;ergs/cm^3$ \\ 
\hline
\end{tabular}
\end{table}
We have chosen $\gamma_{min}=40$ and $\gamma_{max}=1.0\times10^6$ to extend the observed 
spectrum from radio to VHE energies. The approximate expression for synchrotron, SSC and 
EC fluxes (discussed in earlier section and Appendix \ref{appendix:syn} \& 
\ref{appendix:ssc}) are used to derive the parameters whereas for reproducing the 
observed flux in figure (\ref{fig:fit}) the exact expressions are used and solved 
numerically. 
With the given set of parameters if we consider the
number of cold protons  are equal to the number of non-thermal electrons, then the energy
density due to protons will be $\sim 1.5\times10^{-3}\,ergs/cm^3$. 
This is an order less than
the electron energy density (table (\ref{tab:parameters})) and hence its exclusion will
not affect $B_{eq}$ considerably. 

From the figure (\ref{fig:fit}) one can find that model predicts excessive flux at MeV
energies which is nearly an order more than the highest flux detected from 3C\,279 by
{\it EGRET} during its entire mission (\cite{hartman01}). The flux obtained during 2008 
observations by {\it Fermi} is also much lower. Hence it can be argued that this prediction
may be unlikely. However there were no simultaneous observations of source at MeV-GeV
energies during the WEBT campaign and MAGIC detection and hence such a dramatic flare can
not be ruled out. 

The deduced temperature of the dusty torus ($T = 860 K$) is consistent with the 
temperature obtained from the inner region of the dusty torus ($>800 K$) of the nearby 
Seyfert type 2 AGN NGC 1068(\cite{jaffe04}). It is also close to the temperature range
($900-1300 K$) suggested by \cite{devires98} based on near-infrared modelling of
gigahertz peaked spectrum (GPS), compact steep spectrum (CSS) and Fanaroff \& Riley type
II (FR II) sources. 

The emitted VHE photons can be absorbed by soft target photons through photon-photon
pair production mechanism. The condition
on photon energies to be transparent against pair formation is 
\begin{align}
(1+z)(h\nu_\gamma)(h\nu^\star)<(mc^2)^2
\end{align}
If $\nu_{VHE}\approx 10^{26} Hz$ then it can pair produce with target photons with frequency
$\nu^\star>10^{14} Hz$. Hence the emitted VHE photons can pair produce with the BLR 
photons, however the infra red photons from the dusty torus will be below the threshold.
If the emission region is within the BLR region, the VHE photons can
be absorbed by the BLR photons through pair production
(\cite{boettcher09,bai09}). Hence the detection of 3C\,279 at VHE gamma ray energies
demand the emission region may be ahead of the BLR region to avoid the severe $\gamma \gamma$
absorption. \cite{bai09} studied the $\gamma \gamma$ absorption 
of VHE gamma rays from 3C\,279 and suggested 
the emission region must be located within the BLR region. 
However the present work (and also \cite{boettcher09}) require the emission 
region to be far from the BLR 
region to reproduce the simultaneous broadband spectrum of 3C\,279. 

If we consider the dusty torus as an annular ring covering the central
source (\cite{pier92}) then the distance of the inner wall of the torus ($R_{IR}$)
can be estimated as
\begin{align}
R_{IR} &\approx \frac{1}{T^2}\sqrt{\frac{L_{uv}}{4\pi\sigma}} \nonumber \\
\label{eq:torusdis}
&\approx 1.6 \left(\frac{T}{860 K}\right)^{-2}
\left(\frac{L_{uv}}{10^{46} ergs\;s^{-1}}\right)^{1/2} \; parsec
\end{align}
where $L_{uv}$ is the accretion disk luminosity.
Also, if the distance
between BLR clouds and the central source ($R_{BLR}$) of 3C\,279 satisfy the size-luminosity 
relation of \cite{kaspi07}, then
\begin{equation}
R_{BLR} \sim 0.052 \left(\frac{\lambda L_{\lambda}(1350 \;{\rm \AA})}
{10^{46} ergs\; s^{-1}}\right)^{0.52} \; parsec
\end{equation}
For the considered variability timescale of $1$ day (\cite{hartman96}), the distance of the 
emission region from the central source ($R_{fl}$) can be approximated as
\begin{equation}
R_{fl} \sim c t_{var} \Gamma^2\approx 0.6 \left(\frac{t_{var}}{1 day}\right)
\left(\frac{\Gamma}{24}\right)^2 parsec
\end{equation}
which is $\approx 10$ times farther than $R_{BLR}$ supporting our inference. Also 
from equation(\ref{eq:torusdis}) the emission region lies within the dusty torus
i.e $R_{BLR}< R_{fl}< R_{IR}$.

\cite{boettcher09} suggested a multi-zone leptonic model as a 
possibility to explain the VHE emission from 
3C\,279 since one zone model require low magnetic fields and/or very high Lorentz factor of the jet. 
However they arrived to this conclusion by considering the BLR photons as the target photons 
for the external 
Compton process instead of infrared photons from the dusty torus. 
\cite{blazejowski00} considered a model which is 
similar to the one described in the present work, but their aim was to project the importance of 
the Comptonisation of infrared photons from the torus to reproduce the high energy spectrum. 
In the present work, we have used simultaneous observation of 3C\,279 at
different energies to
deduce the physical parameters of the source.

\section{Conclusion}

We reproduce the observed simultaneous broadband spectrum of 3C\,279 using a simple one zone leptonic
model considering synchrotron, SSC and EC processes. From the radio/optical synchrotron spectrum we 
show that the VHE emission cannot be attributed to SSC process whereas EC scattering of IR photons 
from the dusty torus can explain the observed VHE emission. Interpreting the X-ray emission as 
continuation of EC spectrum require magnetic field much lower than its equipartition value.
However an explanation based on SSC origin of X-ray require a magnetic field which is comparable
to the equipartition value.

The model predicts large flux at MeV-GeV energies. Since the data in these 
energy range is not available during the considered flare, such a prediction cannot be ruled out. 
The model parameters describing the source are estimated by considering a magnetic field which
is in equipartition with the particle energy density. A deviation 
from this condition will reflect in the value of the estimated parameters. 
Also if the jet matter contains considerable amount of energetic protons, then the contribution
from these protons should be included in equipartition magnetic field.
However in the present work, the proton contribution to the total particle energy is 
considered to be negligible. Also the size of the emission region estimated from equation(\ref{eq:size}) 
is only an upper 
limit and the actual size may be smaller. Moreover in reality the size of the emission region may
be different for different energy bands. These variation in the emission region size will also be
reflected in the estimated parameters. 

\section{Acknowledgements}
SS thanks Abhas Mitra for constant support and encouragement.  
SG acknowledge Stephan LeBohec and David Kieda for constant support and encouragement and 
the financial support by the National Science Foundation grant PHY 0856411 and PHY 0555451.
Authors also thank Werner Collmar for providing the ASCII-data for {\emph SWIFT}, {\emph Chandra}
and {\emph INTEGRAL} observations.
This research has made use of the NASA/IPAC Extragalactic Database (NED) which is operated 
by the Jet Propulsion Laboratory, California Institute of Technology, under contract with 
the National Aeronautics and Space Administration.

\appendix
\section[]{Synchrotron Emissivity}{\label{appendix:syn}}
The synchrotron emissivity $\epsilon_{syn}(\nu)$ at frequency 
$\nu$ due to a relativistic electron distribution $N(\gamma)$ can be written as
\begin{equation}
\label{eq:appasyn}
\epsilon_{syn}(\nu) = \frac{1}{4\pi}\int\limits_1^\infty P_{syn}(\gamma,\nu)N(\gamma)d\gamma
\end{equation}
Where $P_{syn}(\gamma,\nu)$ is the average synchrotron power emitted by an electron 
with energy $\gamma m c^2$ at frequency $\nu$. Following \cite{shu91}, we can express
$P_{syn}(\gamma,\nu)$ as
\begin{equation}
P_{syn}(\gamma,\nu) = \frac{4}{3}\beta^2\gamma^2c\sigma_TU_B\phi_\nu(\gamma) 
\end{equation}
where $\beta c(\approx c)$ is the velocity of the electron and the function $\phi_\nu(\gamma)$ satisfies
\begin{equation}
\int\limits_0^\infty \phi_\nu(\gamma) d\nu = 1
\end{equation}
If we approximate $\phi_\nu(\gamma)$ as a delta function
\begin{equation}
\phi_\nu(\gamma) \to \delta(\nu-\gamma^2\nu_L)
\end{equation}
Then we can perform the integration (\ref{eq:appasyn}) and the synchrotron emissivity will be 
\begin{equation}
\label{eq:appaemiss}
\epsilon_{syn}(\nu)\approx\frac{\sigma_TcB^2}{48 \pi^2}\nu_L^{-\frac{3}{2}}
  N\left(\sqrt{\frac{\nu^\prime}{\nu_L}}\right)\nu^{\prime{\frac{1}{2}}}
\end{equation}

\section[]{SSC Emissivity}{\label{appendix:ssc}}
The SSC emissivity $\epsilon_{ssc}(\nu)$ at frequency 
$\nu$ due to a relativistic electron distribution $N(\gamma)$ can be written as
\begin{equation}
\label{eq:appbssc}
\epsilon_{ssc}(\nu) = \frac{1}{4\pi}\int\limits_1^\infty P_{ssc}(\gamma,\nu)N(\gamma)d\gamma
\end{equation}
Where  $P_{ssc}(\gamma,\nu)$ is the average SSC power emitted by an electron
with energy $\gamma m c^2$ at frequency $\nu$ and can be written as
\begin{equation}
P_{ssc}(\gamma,\nu) = \frac{4}{3}\beta^2\gamma^2c\sigma_T\int\limits_0^\infty U(\xi) \; d\xi \;
\psi_\nu(\xi,\gamma)
\end{equation}
where 
\begin{equation}
U_{ph} = \int\limits_0^\infty U(\xi) d\xi
\end{equation} 
is the energy density of the target photons (in this case synchrotron photons) and 
$\xi$ is the frequency of the target photon. The function $\psi_\nu(\xi,\gamma)$
will satisfy the condition
\begin{equation}
\int\limits_0^\infty \psi_\nu(\xi,\gamma) d\nu = 1
\end{equation}
Since the frequency($\nu$) of the photon scattered at Thomson regime is $\nu\approx\gamma^2\xi$,
we can approximate $\psi_\nu(\xi,\gamma)$ as a delta function
\begin{equation}
\psi_\nu(\xi,\gamma) \to \delta(\nu-\gamma^2\xi)
\end{equation}
The SSC emissivity will then be
\begin{equation}
\epsilon_{ssc}(\nu) \approx \frac{1}{3\pi}c\sigma_T\int\limits_1^\infty 
U\left(\frac{\nu}{\gamma^2}\right)N(\gamma) d\gamma
\end{equation}
If we write 
\begin{equation}
U(\xi) = \frac{4\pi R}{c}\epsilon_{syn}(\xi)
\end{equation}
then from equation(\ref{eq:appaemiss}) we get
\begin{equation}
\epsilon_{ssc}(\nu) \approx \frac{R c}{36\pi^2}\sigma_T^2B^2\nu_L^{-\frac{3}{2}}\nu^{\frac{1}{2}}
\int\limits_1^\infty \frac{d\gamma}{\gamma}N\left(\frac{1}{\gamma}\sqrt{\frac{\nu}{\nu_L}}\right)
N(\gamma)
\end{equation}
For a broken power law electron distribution (equation(\ref{eq:broken})) we get
\begin{equation}
\epsilon_{ssc}(\nu) \approx \frac{R c}{36\pi^2}K^2\sigma_T^2B^2\nu_L^{-\frac{3}{2}}
\nu^{\frac{1}{2}}\mathnormal{f}(\nu)
\end{equation}
where 
\begin{align}
\mathnormal{f}(\nu) = \Bigg[\left(\frac{\nu}{\nu_L}\right)^{-\frac{p}{2}}
ln\left(\frac{\gamma_1}{\gamma_2}\right) 
+\frac{\gamma_b^{(q-p)}}{q-p}\left(\frac{\nu}{\nu_L}\right)^{-\frac{q}{2}}
\nonumber \\ \times
(\gamma_1^{(q-p)}-\gamma_{min}^{(q-p)})
\Theta\left(\frac{1}{\gamma_b}\sqrt{\frac{\nu}{\nu_L}}-\gamma_{min}\right)
\Bigg]\Theta(\gamma_2-\gamma_1)
\nonumber \\ 
+\Bigg[\gamma_b^{2(q-p)}\left(\frac{\nu}{\nu_L}\right)^{-\frac{q}{2}}
ln\left(\frac{\gamma_4}{\gamma_3}\right) +
\frac{\gamma_b^{(q-p)}}{q-p}\left(\frac{\nu}{\nu_L}\right)^{-\frac{p}{2}}
\nonumber \\ \times
(\gamma_4^{(p-q)}-\gamma_{max}^{(p-q)})
\Theta\left(\gamma_{max}-\frac{1}{\gamma_b}\sqrt{\frac{\nu}{\nu_L}}\right)\Bigg]
\Theta(\gamma_4-\gamma_3)
\end{align}
here $\Theta$ is the Heaviside function and 
\begin{align}
\gamma_1 = MAX\left(\gamma_{min},\frac{1}{\gamma_b}\sqrt{\frac{\nu}{\nu_L}}\right) \\
\gamma_2 = MIN\left(\gamma_b,\frac{1}{\gamma_{min}}\sqrt{\frac{\nu}{\nu_L}}\right) \\
\gamma_3 = MAX\left(\gamma_b,\frac{1}{\gamma_{max}}\sqrt{\frac{\nu}{\nu_L}}\right) \\
\gamma_2 = MIN\left(\gamma_{max},\frac{1}{\gamma_b}\sqrt{\frac{\nu}{\nu_L}}\right) 
\end{align}

\label{lastpage}

\end{document}